\documentstyle[amssymb,prb,aps]{revtex}

\begin{document}

\twocolumn[\hsize\textwidth\columnwidth\hsize\csname 
@twocolumnfalse\endcsname

\title{Equilibrium basal-plane magnetization of superconductive YNi$_2$B$_2$C - the influence of non-local electrodynamics}
\author{J.R. Thompson$^{1,2}$, A.V. Silhanek$^3$, L. Civale$^3$, K.J. Song$^2$, C.V. Tomy$^4$, and D.McK. Paul$^5$ }

\address{$^1$Oak Ridge National Laboratory, Oak Ridge, Tennessee 37831-6061.\\
$^2$Department of Physics, University of Tennessee, Knoxville, Tennessee 37996-1200.\\
$^3$Comisi\'{o}n Nacional de Energ\'{\i}a At\'{o}mica-Centro At\'{o}mico Bariloche and Instituto Balseiro, 8400 Bariloche, Argentina.\\
$^4$Department of Physics,  I.I.T. Powai, Mumbai, 400076, India \\
$^5$Department of Physics, University of Warwick, Coventry, CV4 7AL, United Kingdom. }

\date{submitted October 9, 2000; revised 22 February 2001}
\maketitle

\begin{abstract}
For a single crystal of YNi$_2$B$_2$C superconductor, the equilibrium magnetization $M$ in the square basal plane has been studied experimentally as a function of temperature and magnetic field.  While the magnetization $M(H)$ deviates from conventional London predictions, a recent extension of London theory (to include effects of non-local electrodynamics) describes the experiments accurately.  The resulting superconductive parameters are well behaved. These results are compared with corresponding findings for the case with $M$ perpendicular to the basal plane.

\end{abstract}

\vskip1pc] \narrowtext

\section{Introduction}

Borocarbide superconductors continue to reveal new and interesting features. 
\cite{canfield-98a} This family of compounds, ${\it Re}$Ni$_{2}$B$_{2}$C,
exhibits superconductivity for ${\it Re}=$ Lu, Y, Tm, Er, Ho and Dy. The
last four of these order antiferromagnetically below the Neel temperature $%
T_{N}$ that ranges from $1.5$ K to $10$ K. The relatively high
superconducting transition temperature $T_{c}$ and the broad variation of
the ratio $T_{N}/T_{c}$ within the family make these materials particularly
appropriate to explore the microscopic coexistence of superconductivity and
localized magnetic moments.\cite
{yaron96,Eisaki-94,Eskildsen-98,Gammel-99,Norgaard-00}

A second remarkable feature in these compounds is the formation of
superconducting vortex lattices [FLL] with symmetries other than the
hexagonal one.\cite{yaron96,yethiraj-97a,dewilde-97a,mckpaul-98a} The
presence of these non-hexagonal lattices has been attributed to the effects
of nonlocal electrodynamics, which arise when the electronic mean free path $%
\ell$ is larger than the BCS zero temperature superconducting coherence
length $\xi_0$. Nonlocal electrodynamics in superconductors had been
traditionally associated with very low values of the Ginzburg Landau
parameter, $\kappa \sim 1$. Borocarbide superconductors have $\kappa$ values
in the range of 10 to 20. However, the availability of very clean single
crystals with large $\ell$ permits the observation of nonlocal effects in
these intermediate $\kappa$ materials.

Nonlocality influences the equilibrium magnetic response of the vortex
lattice in various ways. Song et al.\cite{song-99a} have shown that, when
the applied field ${\bf {H}}$ is parallel to the crystallographic $c$-axis
of a YNi$_2$B$_2$C crystal, the reversible magnetization in the mixed state
deviates from the logarithmic dependence on magnetic field, $M \propto
ln(H_{c2}/H)$, which is expected from the standard London model.\cite
{kogan-88a} Such deviations could be quantitatively accounted for within the
framework of a nonlocal generalization of London theory, as developed by
Kogan and Gurevich.\cite{kogan-96a}

More recently, we have measured\cite{civale-99a} $M$ of this compound for $%
{\bf {H}}$ lying within the basal plane, and we found that it oscillates
with angular periodicity $\pi /2$, a behavior that is incompatible with the
standard London model. Indeed, the familiar superconductive mass anisotropy
with components $m_{ij}$, which plays a major role in the layered high-$%
T_{c} $ materials, is a second rank tensor. Thus, in the square ab basal
plane of this tetragonal compound one has $m_{aa}=m_{bb}$, which immediately
implies that the response within the plane should be isotropic. In contrast,
Kogan's nonlocal scenario \cite{kogan-96a} contains a fourth rank tensor
that breaks the basal plane isotropy and provides for the observed four-fold
anisotropy.

In this paper we expand the analysis of the reversible magnetization of the
vortex system in YNi$_2$B$_2$C by presenting detailed measurements of $%
M(H,T) $ for ${\bf {H}}$ in the basal plane of the crystal. Within the
framework of the Kogan-Gurevich model\cite{kogan-96a}, we obtain explicit
expressions for the magnetization in the ab-plane by expanding the free
energy appropriate for this configuration\cite{kogan-99a} to first order in
the basal-plane anisotropy. This approximation provides an excellent
description of the experimental results and, furthermore, the resulting
parameters for the superconductor are well behaved and exhibit a remarkable
consistency with results from band structure calculations. These aggregate
findings give unimpeachable evidence for a profound impact of nonlocal
electrodynamics on clean, intermediate $\kappa$ borocarbide superconductors.

\section{Theoretical background}

Standard local London anisotropic theory provides a simple logarithmic field
dependence for the equilibrium magnetization. For intermediate fields $%
H_{c1}\ll H\ll H_{c2}$, one has\cite{kogan-88a} (for ${\bf H}$ parallel to
the $k$-th principal axis)

\begin{equation}
\frac{M^{k}}{M_{0}^{k}}=-ln\left( \frac{\eta H_{c2}^{k}}{B}\right) ,\hspace{%
0.4in}M_{0}^{k}=\frac{\Phi _{0}}{32\pi ^{2}\lambda _{i}\lambda _{j}}
\label{london}
\end{equation}

Here $\eta $ is a constant of order unity; $\lambda _{i}$ is the London
penetration depth corresponding to screening by currents in the ${\bf j}$%
-direction, with ${\bf {H}\parallel {k}}$-axis; $H_{c2}^{k}=\Phi _{0}/2\pi
\xi _{i}\xi _{j}$ is the upper critical field in the ${\bf k}$-direction,
with $\xi $ being the Ginzburg-Landau coherence length at temperature $T$. \
Experimentally, we will make the usual approximation that the flux density $%
B=H+4\pi M$ can be replaced by the applied field $H$, since the
magnetization $M\ll H$ in all cases treated here.

In the Kogan-Gurevich non-local formulation of London theory\cite{kogan-96a}
there is a third independent length scale, the nonlocality radius $\rho$,
which depends on $\ell$ and $T$ and also reflects the material anisotropy.
It has the form $\rho = \lambda \sqrt{n}$, where $\lambda = \left(
\lambda_a\lambda_b\lambda_c\right)^{1/3}$ and $n$ is the appropriate
component of the forth rank tensor $\hat{n}$, given by

\begin{equation}  \label{n}
\lambda^2 n_{ijlm} \propto \gamma(T,\ell)\langle v_i v_j v_l v_m
\rangle/\langle v^2 \rangle^2.
\end{equation}

Here ${\bf v}$ is the Fermi velocity and $\langle ...\rangle $ indicates
averages over the Fermi surface. The function $\gamma (T,\ell )$, that
contains all the temperature and mean free path dependencies, has been
evalulated by Kogan et al. \cite{kogan-96a,KMP}

When non-locality is important, the scale $H_{c2} \sim \Phi_0/\xi^2$ for $M$
is replaced by another magnetic field scale $H_0 \sim \Phi_0/\rho^2$. As $%
\gamma(T,\ell)$ slowly decreases with increasing $T$, so does $\rho$, and
consequently $H_0$ slowly increases with temperature, in contrast to $H_{c2}$%
. One consequence of the theory is that the quantity $\gamma H_0$ $\sim$ $1/
\xi_0^2$ should be independent of temperature, which is a prediction that we
test later.

In a tetragonal material $\hat{n}$ has four independent components, $%
n_1=n_{aaaa}$, $n_2=n_{aabb}$, $n_3=n_{cccc}$ and $n_4=n_{aacc}$. For ${\bf {%
H} \parallel {c}}$-axis, the resulting expression for $M^{c}(H,T)$ is\cite
{kogan-96a}

\begin{equation}
\frac{M^{c}}{M_{0}^{c}}=-ln\left( \frac{H_{0}^{c}}{B}+1\right) -\frac{%
H_{0}^{c}}{\left( H_{0}^{c}+B\right) }+\zeta ^{c}  \label{kogan}
\end{equation}
where $M_{0}^{c}=\Phi _{0}/32\pi ^{2}\lambda _{ab}^{2}$, $H_{0}^{c}=\Phi
_{0}/4\pi ^{2}\lambda ^{2}n_{2}$ (for a square FLL), and $\zeta ^{c}(T)=\eta
_{1}-ln(H_{0}^{c}/\eta _{2}H_{c2}^{c}+1)$, with both $\eta _{1}$ and $\eta
_{2}$ constants of order unity.

If ${\bf {H}}$ lies in the ab-plane the analysis is more complex. In a
previous study,\cite{civale-99a} we described the basal plane anisotropy in $%
M$ by assuming the validity of an expression analogous to Eq. (\ref{kogan})
and proposing a four-fold oscillation in $H_0$. The empirical expression so
obtained successfully captured the basic features of the in-plane magnetic
response, but the link between the amplitude of the oscillations and the
more fundamental material parameters was undefined.

According to a subsequent generalization of the formalism, developed\cite
{kogan-99a} for the case of ${\bf {H}}$ in the ab-plane, the free energy is
approximately given by

\begin{equation}  \label{integralFab}
F=M_0^{ab} B \int_{u_1}^{u_2} \frac {du}{\sqrt{(u+n_4)(u+n_4+d(\varphi)/4)} }
\end{equation}

Here $M_{0}^{ab}=\Phi _{0}/32\pi ^{2}\lambda _{ab}\lambda _{c}$, $%
u_{1}=(4\pi ^{2}\kappa ^{2})^{-1}$ and $u_{2}=2\pi u_{1}\left(
H_{c2}/B\right) $. (Within this approximation, the very small anisotropy\cite
{Johnson-Halperin} in $H_{c2}$ can be ignored.) Equation (\ref{integralFab})
holds whenever $u\ll 1$, so that terms of order $u^{2}$ can be neglected.
For YNi$_{2}$B$_{2}$C we have $\kappa \sim 10$, thus $u_{1}\sim 2.5\times
10^{-4}$ and $u_{2}\sim 1.6\times 10^{-3}H_{c2}/B$. In our analysis of the
basal-plane magnetization, $\left( H_{c2}/B\right) \leq 40$ in all cases
(see section IV) thus $u_{2}\leq 0.05$ and the approximation is valid.

The in-plane anisotropy is accounted for by $d(\varphi)$,

\begin{equation}  \label{d}
d=n_3 \Gamma^2+ \frac {n_1}{\Gamma^2}-6n_4-\frac {n_1 -3n_2}{2\Gamma^2}
sin^2(2\varphi)
\end{equation}
where $\Gamma^2=m_c/m_a$ is the usual mass anisotropy between the $a$ and $c$%
-axes, and $\varphi$ is the angle between the vortices and the $a$-axis. If $%
d=0$ the in-plane response would be isotropic and integration of Eq. (\ref
{integralFab}) would result in a magnetization $M = -\partial{F}/\partial{B}$
identical to Eq. (\ref{kogan}). However, for $d \neq 0$ the in-plane
magnetization cannot in general be reduced\cite{kogan-99a} to the form (\ref
{kogan}).

The integrand of (\ref{integralFab}) can be expanded in powers of the
variable $d(\varphi)/4(u+n_4)$ and integrated term by term. The resulting
series can be differentiated with respect to $B$ to obtain $M^{ab}$. The
leading term, $M_{iso}^{ab}$, is independent of $d$ and represents a large
contribution to the magnetization that is isotropic within the plane

\begin{equation}
\frac{M_{iso}^{ab}}{M_{0}^{ab}}=-ln\left( \frac{H_{0}^{ab}}{B}+1\right) -%
\frac{H_{0}^{ab}}{\left( H_{0}^{ab}+B\right) }+\zeta ^{ab}  \label{Miso}
\end{equation}
where $H_{0}^{ab}=\Phi _{0}/4\pi ^{2}\lambda ^{2}n_{4}$ (Also in this case,
the numerical factors are valid for a square FLL). The term $\zeta ^{ab}$
arises from the core contribution to the free energy (not included in Eq. (%
\ref{integralFab})); it is analogous to $\zeta ^{c}$ in both form and
origin, as described previously.\cite{kogan-96a,kogan-99a} Eq. (\ref{Miso})
has the same functional form as Eq. (\ref{kogan}), but the anisotropy
between the $c$-axis and the plane is reflected in both the prefactor $M_{0}$
and the field scale $H_{0}$.

The second term in the expansion, $M_1^{ab}$, is linear in $d(\varphi )$ and
accounts for the in-plane fourfold anisotropy in $M^{ab}$ to leading order:

\begin{equation}  \label{M1}
\frac {M_1^{ab}}{M_0^{ab}}=- \frac {d}{8n_4} \frac {B}{B+H_0^{ab}} \left[ 
\frac {H_0^{ab}}{H_0^{ab}+B}- \frac {\frac {H_{c2}}{B}-1}{(\frac {H_{c2}}{%
H_0^{ab}}+1)} \right]
\end{equation}

In the above scenario $H_{c2}$ tends toward zero as $T \rightarrow T_{c}$,
while $H_{0}$ increases with temperature. These differing temperature
dependencies mean that for clean materials, the non-local expressions Eqs. 
\ref{kogan} and \ref{Miso} reduce to the local form, Eq. \ref{london}, as $T$
approaches $T_{c}$, while $M_1^{ab}$ from Eq. \ref{M1} vanishes. Thus, the
non-local theory predicts that the equilibrium magnetization in a clean
sample should vary logarithmically with field near $T_{c}$, but deviate
progressively from logarithmic behavior at low temperatures. Furthermore, as
materials become dirtier so that $\rho $ becomes shorter and $H_{0}$
increases, the non-local expressions reduce to the local form {\it at all
temperatures}.

Our purpose is to analyze the in-plane magnetization using Eqs. \ref{Miso}
and \ref{M1}. We thus need to estimate the importance of the various terms
in the expansion. According to electron band calculations,\cite{Pickett} for
YNi$_2$B$_2$C we have

\begin{eqnarray}
\langle v^2 \rangle & = & 1.50 \times 10^{15} (cm/sec)^2  \nonumber \\
\langle v_a^2 \rangle & = & 0.87 \times 10^{15} (cm/sec)^2  \nonumber \\
\langle v_c^2 \rangle & = & 0.85 \times 10^{15} (cm/sec)^2  \nonumber \\
\langle v_a^4 \rangle & = & 1.15 \times 10^{30} (cm/sec)^4  \nonumber \\
\langle v_a^2 v_b^2 \rangle & = & 1.75 \times 10^{29} (cm/sec)^4  \nonumber
\\
\langle v_c^4 \rangle & = & 7.71 \times 10^{29} (cm/sec)^4  \nonumber \\
\langle v_a^2 v_c^2 \rangle & = & 2.41 \times 10^{29} (cm/sec)^4
\end{eqnarray}

Based on those values we can calculate the relations between all the
components of $\hat{n}$ using Eq. (\ref{n}). In section IV we will compare
these band calculation estimates with our experimental results. In
particular, we are interested in the dimensionless prefactor $d/8n_4$ that
sets the order of magnitude in Eq. \ref{M1}. It is also useful to split $d$
in two parts, $d=d_1+d_2 \sin^2(2\varphi)$. Using Eq. \ref{d} we obtain

\begin{equation}  \label{estimates}
\frac {d_1}{8n_4} \approx 0.23 \hspace{.5in} \frac {d_2}{8n_4} \approx -0.13
\end{equation}

In section IV we will make use of these estimates to gain an idea of the
goodness of our approximations. For comparison, we can calculate the
equivalent values for the similar material LuNi$_{2}$B$_{2}$C, using the
Fermi-surface averages given previously.\cite{kogan-97} The results are $%
d_{1}/8n_{4}\approx 0.42$ and $d_{2}/8n_{4}\approx -0.20$. The larger values
for the Lu-based system mean that the first-order expansion in Eqs. 6-7 is
more accurate for the yttrium-based compound, while the amplitude of
oscillation in the basal plane magnetization can be larger for LuNi$_{2}$B$%
_{2}$C.

\section{Experimental aspects}

The YNi$_{2}$B$_{2}$C single crystal was grown by a high temperature flux
method using Ni$_{2}$B flux, using isotopic $^{11}$B to reduce neutron
absorption in complementary scattering studies. The 17 mg crystal is the
same as that used in previous investigations by Song et al.\cite{song-99a}
and Civale et al. \cite{civale-99a}  It is a slab of thickness $t\sim 0.5mm$
in the $c$-axis direction, with a mosaic spread of less than $0.2^{\circ }$,
as determined by neutron diffraction. In the basal plane the shape is
approximately elliptical with principal axes of length $\sim 2.0$ and $2.5mm$%
, which approximately coincide with the two equivalent $\langle 110\rangle $
axes of the tetragonal structure.

Magnetic studies were conducted in a SQUID-based magnetometer Quantum Design
MPMS-7 equipped with a compensated 70 kOe magnet. Normally, scan lengths of $%
3cm$ were used. The crystal was glued onto a thin Si-disk and mounted in a
Mylar tube for measurement with the magnetic field ${\bf H}$ applied in the
basal plane, along either the $[100]$ or the $[110]$ axis, with an accuracy
better than $3^{\circ}$. For both orientations, the magnetization ${\bf M}$
is parallel to ${\bf H}$ by symmetry, thus only the longitudinal component
measured by the magnetometer needs to be considered. This is also true with $%
{\bf H} \parallel [001]$, as was the case for the previous data that will be
included in our analysis. Then, from now on we will ignore the vector nature
of ${\bf M}$ and denote it simply as $M$. The diamagnetic moment of the
addenda (silicon disk plus glue), which was measured separately, was linear
in $H$, isotropic, and non-hysteretic. This signal, $m_{Si} = (-5.4 \times
10^{-9} emu/Oe) H$, was always small compared with the moment of the
crystal, and was subtracted from all the data prior to any further analysis.

In the mixed state, hysteresis loops $M(H,T)$ were measured. The maximum $H$
was $65kOe$ in all cases. Measurements were also conducted in the normal
state at temperatures up to $300K$, in order to correct for the normal state
background moment. \ It is worth noting that the magnetization is small,
compared to the applied field, in all cases considered here. \ Thus
demagnetizing efects are negligible: \ for ${\bf H}\parallel ab$-plane, we
obtain from the Meissner state response\cite{civale-99a} that the effective
demagnetization factor $D\thickapprox 0.1$. Then in the mixed state, the
effective field $H_{eff}=H_{applied}-4\pi DM$ differs from the applied field
by less than 1 \% is the worst case and we need to consider only the
magnetizing field ''$H$.'' \ Furthermore, for comparison with theoretical
expressions, we can approximate the flux density $B=H+4\pi M$ by $H$ with an
accuracy of a few percent and generally better.

The superconductive transition temperature, measured in a small applied
field, was $T_{c}=14.5K$. Measurement of the electrical resistivity using a
van der Pauw method gave an electrical resistivity of $4\mu \Omega -cm$ at $%
20K$ and a residual resistance ratio of 10, yielding an electronic mean free
path $\ell \thickapprox 300%
\mathop{\rm \AA }%
$.\ \ Using this and values of $H_{c2}$ (for $H$ $\Vert c$-axis), Song et
al. deduced the values $\xi _{0}=120$ $%
\mathop{\rm \AA }%
$ for the BCS coherence length at $T=0$ and $\kappa \sim 10$. \ Other
superconducting parameters for this compound are collected in Table I.

\section{Results and discussion}

For magnetic field orientations ${\bf H}\parallel \lbrack 100]$ and ${\bf H}%
\parallel \lbrack 110]$, we measured isothermal magnetization loops $M(H)$
at temperatures $T=3$ to $14K$, in $1K$ intervals, and also at $18K$,
slightly above $T_{c}$. Three of those loops, with ${\bf H}\parallel \lbrack
110]$, are shown in Fig. 1, at temperatures $T=5K$, $12K$ and $18K$, i.e.,
well below, near, and just above $T_{c}$. The magnetic response of this
material in the superconducting mixed phase is slightly irreversible,
reflecting weak pinning of flux lines. Now, as the only source of magnetic
hysteresis is vortex pinning, $M(H)$ becomes reversible for $H>H_{c2}(T)$.
Based on the Bean's critical state model, we calculated the equilibrium
magnetization as the average $M_{eq}(H)=\left[ M^{\uparrow }+M^{\downarrow }%
\right] /2$ of the magnetization values measured in the field-increasing and
field-decreasing branches of the loop, respectively. The result for $T=5K$
is shown as a dashed line in Fig. 1. The identification of $M_{eq}(H)$ with
the average between the branches of the loop increases in accuracy as the
width of the hysteresis loop $\left[ M^{\uparrow }-M^{\downarrow }\right] $
decreases. For each temperature, we disregard the low field data where large
hysteresis introduces a significant uncertainty in the determination of $%
M_{eq}$. \ Experimentally, as the field decreases, the width of the $M(H)$
loop increases continuously and smoothly. \ Then at some
temperature-dependent low field, $M$ changes abruptly as it reaches a
minimum and starts to increase. This feature is visible in Fig. 1 for $T$ =
5 and 12 K, for example. Below this field, the hysteresis grows suddenly. We
compute the average magnetization starting at a field slightly above this
minimum.

It is apparent in Fig. 1 that the magnetization $M_{eq}$ does not vanish
above $H_{c2}$ and above $T_{c}$. This indicates the presence of a normal
state contribution $M_{ns}(H,T)$ to the magnetization (recalling that the
signal from the sample holder has been already subtracted). Pure YNi$_{2}$B$%
_{2}$C has no localized moments, thus in the normal state it is expected to
exhibit a linear and nominally temperature-independent paramagnetic (Pauli
and Van-Vleck) susceptibility $\chi _{0}$. However, close inspection of the
data indicates that this intrinsic term cannot account for the entire normal
state signal. Indeed, $M_{ns}(H,T)$ is not linear in $H$ and it grows as $T$
decreases, thus pointing to the presence of localized moments. The small
magnitude of the signal suggests, on the other hand, that this contribution
arises from magnetic impurities. To confirm this, we measured the
temperature dependence of $M_{ns}(H,T)$ from $16K$ to $300K$ at several
fixed fields. The results for $H=1$ and $5kOe$ are shown in the inset of
Fig. 1. As the localized moments are very dilute, no magnetically ordered
phase should appear and one can expect a Curie Law dependence. The solid
lines in the inset are fits to $M_{ns}/H=\left[ \chi _{0}+C/T\right] $. The
Curie term corresponds to a rare earth impurity content of $\sim $ 0.1 at $%
\% $ relative to yttrium, most likely contaminants in the yttrium starting
material.

To isolate the magnetization associated with the vortex state, it is
necessary to remove the paramagnetic background. The normal state
magnetization $M_{ns}$ is well described by the expression

\begin{equation}  \label{brillouin}
M_{ns}=\chi_0 H+M_{sat}B_J \left(\frac{g\mu_B JH}{k_BT}\right)
\end{equation}
where $\chi_0$ is the temperature independent susceptibility obtained as
explained above and $B_J$ is the Brillouin function for effective angular
momentum $J$. For small values of the argument $H/T$, the Brillouin function
leads to the Curie susceptibility, of course.

The magnetic signal arising from the localized moments is clearly observed
in Fig. 2, where $\left[ M_{eq}-\chi _{0}H\right] $ is plotted as a function
of $H/T$. The figure includes experimental data from temperature sweeps at
three values of $H$, and from isothermal loops at temperatures down to $5K$.
At each $T$, the onset of the superconducting signal below $H_{c2}(T)$ is
clearly observed in the nearly vertical trace of data below the envelope
curve.  It is also apparent that for $H>H_{c2}(T)$ all the data at different 
$T$ overlap on a single curve arising from the paramagnetic contribution of
the magnetic impurities. The solid line is a nonlinear fit to Eq. \ref
{brillouin} that yields parameter values $J=3/2$ and $g=6$. The saturation
magnetization $M_{sat}=0.92G$ corresponds to $\sim 7\times 10^{-4}$ per
formula unit content of rare earth impurities, most likely trace
contaminants in the yttrium starting material. \ At this concentration, the
magnetic impurity ions are isolated, but strongly influenced by the crystal
field of the host, producing an anisotropic magnetic response in the normal
state. \ The susceptibility is smaller for $H\Vert c$-axis, as found for \ $%
{\it Re}$Ni$_{2}$B$_{2}$C crystals with \ ${\it Re}$ = Tb, Dy, and Ho, but
not Tm.\cite{Cho}\ \ \ These qualitative similarities show that (a mixture
of) any of several rare earth impurities can generate the observed
paramagnetic response. \ The central and important point here, however, is
that the empirical fit to Eq. \ref{brillouin} provides a precise description
of the normal state signal, so that the superconductive magnetization can be
isolated for analysis.

The resulting (background-corrected) superconducting state equilibrium
magnetization $M=M_{eq}-M_{ns}$ is shown in Fig. 3, as a function of
magnetic field $H$ applied along the $[110]$ axis of the crystal. Results
are shown for temperatures $3K\leq T\leq 14K$ in intervals of $1K$.
Qualitatively, the curves for temperatures near $T_{c}$ are linear, showing
the dependence on $ln(B)$ as predicted by traditional local London theory.
At lower temperatures, however, the increasing curvature visibly signals a
progressive departure from local London behavior. \ 

For a quantitative analysis, we fit the low temperature data to the
non-local relation, Eq. (\ref{Miso}), varying the parameters $M_{0}^{ab}$, $%
H_{0}^{ab}$, and $\zeta ^{ab}$. These fits describe the low temperature
experiments very well, as shown by the solid lines in Fig. 3. The resulting
values of $M_{0}^{ab}$, $H_{0}^{ab}$, and $\zeta ^{ab}$ as a function of $T$
(up to $T=10K$) are shown in Figs. 4(a), (b) and (c) respectively. At higher
temperatures, $T\geq 11K$, the system closely approximates local London
behavior, with $M\propto ln(B)$ and we analyze these data using the local
expression Eq. (\ref{london}). Numerically, the data close to $T_{c}$ have a
very shallow minimum as a function of $H_{0}$ and it becomes meaningless to
fit an (essentially) straight line with three parameters. The fits to the
local London relation are shown in Fig. 3 as dashed lines, and the values of 
$M_{0}(T)$ obtained by this procedure are also included in Fig. 4(a) (open
symbols). In the nonlocal analysis, $H_{c2}$ was not a fitting parameter. It
was determined independently, as explained below.

Before we discuss the results shown in Fig. 4, we must analyze what is the
error involved in taking $M^{ab} \approx M_{iso}^{ab}$, disregarding terms
of higher order. To estimate the contribution $M_1^{ab}$ given by Eq. \ref
{M1}, we must rely on the numerical values (\ref{estimates}) obtained from
band calculations. We see that, if at least the {\it signs} in (\ref
{estimates}) are correct, $M_1^{ab}$ minimizes for $\varphi = 45^{\circ}$.
For that reason we chose to apply Eq. \ref{Miso} to the $[110]$ data and not
to the $[100]$ orientation. Numerically, we have $d(\varphi=45^{%
\circ})/8n_4=\left(d_1+d_2\right)/8n_4 \approx 0.10$. Thus, Eqs. (\ref{Miso}%
) and (\ref{M1}) indicate that $M_1^{ab} \ll M_{iso}^{ab}$ over most of the
field range (for instance, at $T=3K$ and $B=2kG$, $M_1^{ab}/M_{iso}^{ab}
\sim 0.02$, and the same ratio is obtained for 6K and 10kG). We can also
compare the logarithmic field derivatives of $M$, which give the slopes in
Fig. 3. For $B \ll H_0$, to leading order we obtain $\partial
M_{iso}^{ab}/\partial \ln B \approx M_0^{ab}$ and $\partial
M_1^{ab}/\partial \ln B \approx 0.10 \left(2B/H_0\right)M_0^{ab}$, thus for $%
B=2kG$ the error in the slope produced by disregarding the second term in
the expansion is less than $0.5\%$.

The above estimates indicate that by using Eq. (\ref{Miso}) to analyze the
magnetization in the $[110]$ direction we are introducing an error in the
determination of $M_0^{ab}$ and $H_0^{ab}$ of at most a few percent, which
is less than the experimental noise observed in Figs. 4(a) and (b). So, it
is meaningful to discuss the results shown in Fig. 4. For comparison, the
corresponding results for the case\cite{song-99a} with ${\bf H} \parallel
[001]$ are included, too.

Figure 4(a) shows that $M_{0}$ varies linearly with $T$ near $T_{c}$ and
extrapolates to zero at $T_{c}$, consistent with Ginzburg-Landau theory.
From Eq. (\ref{london}), the ratio $M_{0}^{c}/M_{0}^{ab}=\lambda
_{c}/\lambda _{ab}=\sqrt{\left( m_{c}/m_{ab}\right) }=\Gamma $. As expected,
we find that this ratio is rather independent of temperature, $\Gamma
=1.13\pm 0.02$. This result is very comparable with the experimental value
of 1.16 for LuNi$_{2}$B$_{2}$C obtained by Metlushko et al.\cite{Metlushko}
This experimental value can also be compared with the band structure
calculation, since $\Gamma =\sqrt{(\langle v_{a}^{2}\rangle /\langle
v_{c}^{2}\rangle )}$. The resulting value, 1.01, is much smaller than that
deduced experimentally. Returning to the experimentally determined
parameters, we can also calculate $\lambda _{c}(T)$, $\lambda _{ab}(T)$ and $%
\lambda (T)$. The results at $T=3K$ are shown in Table 1.

The next frame, Fig. 4(b), shows $H_{0}(T)$. Qualitatively, $H_{0}$ for both
orientations is constant at low temperature, then increases as $T$
increases, consistent with the theoretically predicted behavior.\cite
{kogan-96a,KMP}  In section II above, we noted that the non-local theory
predicts that the quantity $H_{0}\gamma $ should be independent of
temperature. Previously\cite{song-99a} we found that $\xi _{0}=12$ nm; this
value agrees within experimental error with that calculated from the BCS
expression $\xi _{0}=\hbar \langle v_{F}\rangle /(\pi ^{2}e^{-\gamma
}k_{B}T_{c})$ = 11.6 nm, using $\langle v_{F}\rangle $ = $3.87\times 10^{7}$
cm/sec from band structure,\cite{Pickett} with $e^{\gamma }\approx 1.78$.
Also, from the electrical resistivity, we have that $\ell \approx 30$ nm for
this crystal and therefore evaluate the impurity parameter $\gamma (T)$ with 
$\xi _{0}/\ell =0.3$. The results for $H_{0}\gamma $ are shown as open
symbols in Fig. 4b, for the two orientations of magnetic field. The
constancy of the product provides further solid evidence that non-local
electrodynamics strongly modify the superconductive properties of clean
borocarbides. From the $H_{0}$ data we can extract the nonlocality radius in
both orientations, $\rho _{ab}(T)=\left( \Phi _{0}/4\pi ^{2}H_{0}^{c}\right)
^{1/2}$ (see Song et al.\cite{song-99a}), and analogous expression for $\rho
_{c}(T)$. We can also calculate two of the four independent components of
the tensor $\hat{n}$, namely $n_{2}=\left( \rho _{ab}/\lambda \right) ^{2}$
and $n_{4}=\left( \rho _{c}/\lambda \right) ^{2}$. All the numerical values
for $T=3K$ are listed in Table 1.

Figure 4(c) shows the temperature dependence of the fitting parameter $\zeta$%
. Qualitatively, the data reflects the fact that $\zeta \sim
-\ln\left(H_0/H_{c2}\right)$. As $T$ increases, $H_0/H_{c2}$ varies little
at low temperatures, but then becomes larger, due to the differing
dependencies of $H_0(T)$ and $H_{c2}(T)$, thus $\zeta(T)$ decreases.

The last frame, Fig. 4d, shows $H_{c2}(T)$ in both orientations, obtained
(a) by extrapolating the isothermal magnetization in the superconductive
state to $M = 0$ and (b) by locating the field at which the magnetic
hysteresis disappears. These results were introduced as fixed parameters in
Eq. (\ref{Miso}), as already mentioned. It is evident that $H_{c2}(T)$
curves upward near $T_c$. This differs from simple Ginzburg-Landau behavior,
but is consistent with other observations on this material.\cite
{Shulga,Michor} Also, little anisotropy is evident between the c-axis and
the basal plane values. The mass anisotropy obtained from the $M_0$ data
(Fig. 4(a)) would imply an experimentally resolvable difference in $H_{c2}$.
Previously, Johnson-Halperin et al. \cite{Johnson-Halperin} also found a
nearly isotropic response in single crystal YNi$_2$B$_2$C. The reason for
the discrepancy between the anisotropy derived from $M_0$ and $H_{c2}$
remains unclear. Interestingly, LuNi$_2$B$_2$C, an isostructural,
nonmagnetic borocarbide superconductor with similar $T_c$, exhibits greater
anisotropy in $H_{c2}$, both from the c-axis and within the basal plane.\cite
{Metlushko}

Next we consider the anisotropy of the magnetization within the basal plane.
We have previously shown\cite{civale-99a} that this quantity exhibits a
four-fold periodicity, which cannot be accounted for within the local London
theory. In Fig. 5 we show the amplitude of the oscillation in basal plane
magnetization, $\delta M=M[110]-M[100]$, as a function of field $H$. The
experimental data for $T=7K$ are reproduced from Civale et al.\cite
{civale-99a}, where details of the experimental procedure and data analysis
are provided. Here we analyze these data using Eq. (\ref{M1}), where only
the second term, proportional to $d_{2}$, contributes to the oscillatory
behavior. In this expression, we have already values for $H_{c2}$ and $%
H_{0}^{ab}$, so there is a single unknown, $d_{2}/8n_{4}$, available for
fitting the data. The result is shown as a solid line in Fig. 5. The value
of the sole fitting parameter is $d_{2}/8n_{4}=-0.136$, which we discuss
below.

Let us now summarize the experimentally determined parameters related to the
material anisotropy. Theoretically, five independent parameters ($n_1, n_2,
n_3, n_4$ and $\Gamma$) are necessary to describe fully the angularly
dependencies. Experimentally we have obtained three of these quantities, $%
n_2 $, $n_4$ and $\Gamma$, plus the combination $d_2/8n_4$ that allows us to
obtain the value for $n_1$ using Eq. 5. Thus, we have obtained four out of
the five independent parameters, as shown in Table I. For comparison with
band structure calculations and to eliminate poorly known prefactors, it is
convenient to consider ratios relative to $n_4$, as shown in Table II. The
experimental and calculated ratios agree within $\sim$ 15 $\%$, which is
quite reasonable given the approximations. Lastly we consider the sole
fitting parameter $d_2/8n_4 \sim -0.14$ that determines the amplitude of the
oscillations of $M$ in the basal plane. This quantity has the correct sign
and it lies in remarkable numerical agreement with the band structure value,
-0.13. The excellent description of the data, using parameter values
measured independently here ($M_0$, $H_{c2}$, and $H_0$) and with $d_2/8n_4$
almost coinciding with the calculated value, gives considerable confidence
in the fundamental correctness of the non-local description.

\section{conclusions}

We have measured the equilibrium magnetization in the basal plane of clean,
superconducting YNi$_{2}$B$_{2}$C, taking careful account of background
effects. Influences of non-local electrodynamics are very evident well below 
$T_{c}$, with the effects becoming washed out at higher temperatures as
expected. The magnetic-field-dependence of the magnetization and its
oscillation amplitude are well described by a generalization of London
theory to include non-local influences. The results of this analysis have
been compared with a corresponding study of the c-axis magnetization,
thereby providing the value $\Gamma $ = 1.13 for the mass anisotropy of this
material. Furthermore, the experimental values for the material parameters
agree overall very well with those deduced from the band structure. In
summary, these clean, non-magnetic borocarbide superconductors display a
rich variety of physical phenomena, requiring both 2-nd and especially 4-th
rank tensors to describe their macroscopic vortex state properties.

\section{acknowledgements}

We wish to acknowledge useful discussions with V.G. Kogan, M. Yethiraj and
D. K. Christen and to thank W.E. Pickett for his unpublished band structure
results. A portion of the work of JRT was supported by the Science Alliance
at The University of Tennessee, Knoxville. Research at ORNL was sponsored by
the U.S. Department of Energy under contract DE-AC05-00OR22725 with the Oak
Ridge National Laboratory, managed by UT-Battelle, LLC. Research at CAB-IB
was partially supported by ANPCyT, Argentina, PICT 97 No. 01120, and by
CONICET, Argentina, PIP No. 4207. A.V.S. would like to thank the CONICET for
financial support.

\newpage 
\begin{table}[ht]
\caption{Superconducting parameters}\centering 
\begin{tabular}[b]{lcc}
$\lambda=950\AA$ & $\Gamma=1.13 \pm 0.02$ &  \\ 
$\lambda_{ab}=990\AA$ & $\lambda_{c}=880\AA$ &  \\ 
$\rho_{ab}=31.4\AA$ & $\rho_c=34.4\AA$ &  \\ 
$n_1=6.95.10^{-3}$ & $n_2=1.04.10^{-3}$ &  \\ 
$n_3=-$ & $n_4=1.25.10^{-3}$ & 
\end{tabular}
\end{table}

\begin{table}[ht]
\caption{Average Fermi velocities}\centering 
\begin{tabular}[b]{lcc}
& exp. & band calc. \\ \hline
$n_1/n_4$ & 5.56 & 4.78 \\ \hline
$n_2/n_4$ & 0.83 & 0.73 \\ \hline
$n_3/n_4$ & -- & 3.2
\end{tabular}
\end{table}

{\bf Figure captions}

Fig. 1. The magnetization of single crystal YNi$_{2}$B$_{2}$C at three
temperatures, with magnetic field in the basal plane, ${\bf H} \parallel$
[110] axis. The figure illustrates the limited irreversibility in the
superconductive state below $T_{c} = 14.5 K$, and a paramagnetic background
in the normal state. Inset: Curie analysis of the normal state
susceptibility in the temperature range $T = 18 - 300 K$.

Fig. 2. Total equilibrium magnetization $M_{eq}$ minus $\chi _{0}H$ (see
text) versus $H/T$ obtained from isothermal field sweeps at several $T$ and
from temperature sweeps at three values of $H$, as indicated in the figure.
The solid line is a Brillouin function (Eq. \ref{brillouin}) fitted to the
paramagnetic impurity-background data. The nearly vertical traces of data
show the entry into the superconductive state.

Fig. 3. The equilibrium magnetization $M_{eq}$ in the superconductive state
with $H \parallel $ [110] axis, plotted versus $H$ (logarithmic axis).
Discrete symbols are data measured at temperatures of 14, 13, 12, ..., 3 K;
straight (dashed) lines show conventional, local London behavior near $T_{c}$%
. A nonlocal generalization of London theory (Eq. \ref{Miso}, solid lines)
accounts well for the pronounced deviations from local behavior at lower
temperatures.

Fig. 4. Superconductive parameters of YNi$_{2}$B$_{2}$C with magnetic field $%
H$ $\Vert \lbrack 001]$ or $H$ $\Vert \lbrack 110]$. (a) The magnitude of
the equilibrium magnetization $M_{0}$, where solid symbols denote results
from the non-local analysis and open symbols come from a conventional London
analysis. (b) The field scale $H_{0}$ and the (theoretically constant)
quantity $H_{0}\gamma $ (where $\gamma $ = impurity parameter; see text).
(c) The fitting parameter $\zeta $. (d) Estimates of the upper critical
field $H_{c2}$.

Fig. 5. Amplitude of the oscillation in basal plane magnetization, $\delta
M=M[110]-M[100]$, plotted versus $H$ on a logarithmic axis. Solid line shows
a fit to the oscillatory term in Eq. \ref{M1} , with $H_{0}^{ab}$ = 48 kOe
and $H_{c2}^{ab}$ = 33 kOe; see text.



\section{References}

\begin{references}
\bibitem{canfield-98a}  P.C. Canfield, P.L. Gammel and D.J. Bishop, {\it %
Physics Today} {\bf 51}, 40 (1998).

\bibitem{yaron96}  U. Yaron, P.L. Gammel, A.P. Ramirez, D.A. Huse, D.J.
Bishop, A.I. Goldman, C. Stassis, P.C. Canfield, K. Mortensen, and M.R.
Eskildsen, {\it Nature} {\bf 382}, 236 (1996).

\bibitem{Eisaki-94}  H. Eisaki, H. Takagi, R. J. Cava, B. Batlogg, J. J.
Krajewski, W. F. Peck, Jr., K. Mizuhashi, J. O. Lee, and S. Uchida, {\it %
Phys. Rev. B} {\bf 50}, 647 (1994).

\bibitem{Eskildsen-98}  M. R. Eskildsen; K. Harada; P. L. Gammel; A. B.
Abrahamsen; N. H. Andersen; G. Ernst; A. P. Ramirez; D. J. Bishop, K.
Mortensen, D. G. Naugle, K. D. D. Rathnayaka, and P. C. Canfield, {\it Nature%
} {\bf 393}, 242 (1998).

\bibitem{Gammel-99}  P. L. Gammel, B. P. Barber, A. P. Ramirez, C. M. Varma,
D. J. Bishop, P. C. Canfield V. G. Kogan, M. R. Eskildsen, N. H. Andersen,
K. Mortensen, and K. Harada, {\it Phys. Rev. Lett.} {\bf 82}, 1756 (1999).

\bibitem{Norgaard-00}  K. N\o rgaard, M. R. Eskildsen, and N. H. Andersen,
J. Jensen, P. Hedeg\aa rd, and S. N. Klausen, and P. C. Canfield, {\it Phys.
Rev. Lett.} {\bf 84}, 4982 (2000).

\bibitem{yethiraj-97a}  M. Yethiraj, D.McK. Paul, C.V. Tomy and E.M. Forgan, 
{\it Phys. Rev. Lett.} {\bf 78}, 4849 (1997).

\bibitem{dewilde-97a}  Y. De Wilde, M. Iavarone, U. Welp, V. Metlushko, A.E.
Koshelev, I. Aranson, G.W. Crabtree and P.C. Canfield, {\it Phys. Rev. Lett.}
{\bf 78}, 4273 (1997).

\bibitem{mckpaul-98a}  D.McK. Paul, C.V. Tomy, C.M. Aegerter, R. Cubitt,
S.H. Lloyd, E.M. Forgan, S.L. Lee and M. Yethiraj, {\it Phys. Rev. Lett.} 
{\bf 80}, 1517 (1998).

\bibitem{song-99a}  K.J. Song, J.R. Thompson, M.Yethiraj, D.K. Christen,
C.V. Tomy and D. McK. Paul, {\it Phys. Rev. B} {\bf 59}, R6620 (1999).

\bibitem{kogan-88a}  V.G. Kogan, M.M. Fang, and S. Mitra, {\it Phys. Rev. B} 
{\bf 38}, 11958 (1988).

\bibitem{kogan-96a}  V.G. Kogan, A. Gurevich, J.H. Cho, D.C. Johnston, Ming
Xu, J.R. Thompson and A. Martynovich, {\it Phys. Rev. B} {\bf 54}, 12386
(1996).

\bibitem{civale-99a}  L. Civale, A.V. Silhanek, J.R. Thompson, K.J. Song,
C.V. Tomy and D.McK. Paul, {\it Phys. Rev. Lett.} {\bf 83}, 3920 (1999).

\bibitem{kogan-99a}  V.G. Kogan, S.L. Bud\'{}ko, P.C. Canfield and P.
Miranovic, {\it Phys. Rev. B} {\bf 60}, R12577 (1999).

\bibitem{KMP}  V. G. Kogan, P. Miranovic, and D. McK. Paul, in {\it The
Superconducting State in Magnetic Fields: Special Topics and Trends}, edited
by C. A. R. Sa de Melo, Directions in Condensed Matter Physics, vol {\bf 13}
(World Scientific, Singapore, 1998).

\bibitem{Johnson-Halperin}  E. Johnston-Halperin, J. Fiedler, D. E. Farrell,
Ming Xu, B. K. Cho, P. C. Canfield, D. K. Finnemore, and D. C. Johnston, 
{\it Phys. Rev. B} {\bf 51}, 12852 (1995).

\bibitem{Pickett}  W.E. Pickett (unpublished).

\bibitem{kogan-97}  V.G. Kogan, M. Bullock, B. Harmon, P. Miranovi\'{c}, Lj.
Dobrosavljevi\'{c}-Gruji\'{c}, P.L. Gammel, and D.J. Bishop, {\it Phys. Rev.
B} {\bf 55}, R8693 (1997).

\bibitem{Metlushko}  V. Metlushko, U. Welp, A. Koshelev, I. Aranson, G. W.
Crabtree, and P. C. Canfield, {\it Phys. Rev. Lett.} {\bf 79}, 1738 (1997).

\bibitem{Cho}  B. K. Cho, {\it Physica C} {\bf 298}, 305 (1998).

\bibitem{Shulga}  S. V. Shulga, S. -L. Drechsler, G. Fuchs, K. H. Mueller,
K. Winzer, M. Heinecke, and K. Krug, {\it Phys. Rev. Lett.} {\bf 80}, 1730
(1998).

\bibitem{Michor}  G. Hilscher and H. Michor in {\it Studies of High
Temperature Superconductors}, vol. {\bf 28}, edited by A. V. Narlikar (Nova,
New York, 1999), pp. 28-74 and references therein.
\end{references}
\end{document}